# Awakening catalytically active surface of BaRuO$_3$ thin film for alkaline hydrogen evolution

Jegon Lee[a,1], Do-Hyun Kim[a,1], Seulgi Ji[b,1], Sangmoon Yoon[c], Seung Hyun Nam[a], Jucheol Park[d], Jin Young Oh[a], Seung Gyo Jeong[a], Jong-Seong Bae[e], Sang A Lee[f,*], Heechae Choi[b,g,*], Woo Seok Choi[a,*]

[a]Department of Physics, Sungkyunkwan University, Suwon 16419, Republic of Korea

[b]Institute of Inorganic and Materials Chemistry, University of Cologne, Greinstr. 6, 50939, Cologne, Germany

[c]Department of Physics and Semiconductor Science, Gachon University, Seongnam, 13120 Republic of Korea

[d]Gyeongbuk Science & Technology Promotion Center, Gumi Electronics & Information Technology Research Institute, Gumi 39171, Republic of Korea

[e]Yeongnam Regional Center, Korea Basic Science Institute, Busan 46742, Republic of Korea

[f]Department of Physics, Pukyong National University, Busan 48513, Republic of Korea

[g]Advanced Materials Research Center (AMRC) & Department of Chemistry and Materials Science, School of Science, Xi'an Jiaotong-Liverpool University, Suzhou, 215123, China

* Corresponding authors.
  E-mail: sangalee@gmail.com (S.A Lee), heechae.choi@xjtlu.edu.cn (H. Choi), choiws@skku.edu (W.S. Choi)

[1] Equally contributed to this work.

**Abstract**

The dynamic reconstruction of surfaces during electrochemical reactions plays a crucial role in determining the performance of electrocatalysts. However, because reconstructions occur at the atomic level, direct observation and elucidation of the underlying mechanism are challenging for conventional powder-type catalysts with ill-defined lattices. In this study, the catalytically active surface of 3C $BaRuO_3$ (BRO) epitaxial thin films emerges upon the dynamic introduction of surface Ru clusters, for the alkaline hydrogen evolution reaction (HER). Based on the mass activity at overpotential 100 mV, the intrinsic HER performance increases dramatically from 0.11 to 7.72 A $mg_{Ru}^{-1}$ immediately after the initial HER cycle and eventually saturates at 1.05 A $mg_{Ru}^{-1}$ after continuous operation. The formation of Ru clusters on the catalyst surface, driven by selective Ba leaching under alkaline HER conditions, is observed experimentally. Density functional theory calculations demonstrate that HER activity increased with enhanced H* adsorption owing to the dynamic $Ru_6$ cluster formation. A strategy for stabilizing the 'awakened' active surface of BRO is further proposed by validating that the atomic-scale control of the film thickness can effectively maintain the highly active state. This study offers fundamental insights into the design and stabilization of the highly active Ru-based electrocatalysts for the alkaline HER.

# 1. Introduction

The efficiency of water electrolysis, a critical technology for hydrogen production without carbon emission, is determined by the dynamic surface chemistry of the catalyst [1-6]. The catalytic mechanisms for hydrogen production, including the hydrogen and oxygen evolution reactions (HER and OER, respectively), involve repeated dynamic formation and breakage of chemical bonds between the adsorbate molecules and electrochemically active sites on the catalyst surface [7-9]. Accordingly, efforts to improve the performance of electrocatalysts have focused on enhancing the catalytically active sites primarily located on the surfaces [10-12]. Reinforcing metal-support interactions or constructing core-shell structures by combining a highly active material with a stable one are conventional methods of activating the catalyst surface [13-17]. For example, a highly active but unstable Ru catalyst can be stabilized by fabricating a core-shell structure using a stable Ni core, where the core-shell composite exhibits excellent activity in both the HER and OER [14]. Likewise, Ru nanoparticles can be stabilized by dispersing them onto O-doped graphene decorated with Co atoms, leading to outstanding HER performance [15]. More recently, CoCe metal–organic frameworks with lattice distortion and large specific surface area enable enhanced catalytic activity and long-term stability for practical devices in alkaline HER [17]. However, these approaches only allow tuning of the initial state of the catalyst surface, whereas the surface state of an electrocatalyst is often dynamically reconstructed by repeated electrocatalytic operations [18-22].

Dynamic self-reconstruction of the surface of perovskite oxides via electrocatalytic reactions has been reported to improve the electrocatalytic activity [23-25]. Surface self-reconstruction mainly occurs through leaching and amorphization of catalyst surface due to its thermodynamic instability in electrochemical environments [26-29]. In particular, the atomic dissolution of $A$-site ions in perovskite oxides with an $AB$O$_3$ configuration ($A$: alkaline (earth) or rare-earth metals, $B$: transition metals) has been suggested as one of the major triggers for surface self-reconstruction. Recently, enhancement of the HER performance by self-reconstruction has been reported for bulk strontium ruthenate catalysts.

Sr leaching during the electrochemical cycles led to the slow formation of $RuO_2$ and Ru nanoparticles on the surfaces of $SrRuO_3$ and $Sr_2RuO_4$, respectively [28-30]. Because this electrochemical self-reconstruction occurs at a nanometer-scale depth from the nominal surface, an atomic-scale-precision engineering of a material system is necessary to identify the reconstruction mechanism.

Epitaxial thin film geometry by atomic-scale-precision growth enables unique identification of self-reconstruction of the catalyst surface during the electrocatalytic reactions, which would not have been possible using typical powder-type bulk catalysts. The overall efficiency of the thin film catalyst might be inferior to that of congeners with powder-type geometries with maximized surface-to-volume ratios. Nevertheless, it offers a unique advantage. Using an epitaxial thin film geometry allows us to gain fundamental insight into the intrinsic electrocatalytic mechanisms, such as self-reconstruction, which is otherwise inaccessible using bulk systems. For example, electrochemical activation via the formation of a highly active $IrO_x$ surface phase has been reported using epitaxial $SrIrO_3$ thin films [29], and the electrocatalytic contribution of a subsurface layer as deep as 10 perovskite unit cells has been reported using heterostructures with atomic-scale-precision epitaxy [31]. Moreover, by using a thin film system, tuning parameters such as the epitaxial strain [32-36], lattice symmetry [37, 38], crystal facets controlled by the substrate orientations [39-41], and elemental defects/defect clusters [42-45], can be effectively isolated to identify the intrinsic mechanisms of electrocatalytic activities.

Herein, we demonstrate that a highly performing electrocatalytic surface for alkaline HER can be activated and saturated in a $BaRuO_3$ (BRO) epitaxial thin film catalyst via self-reconstruction. Among perovskite Ruthenates, BRO was chosen as it shows superior HER performance and ideal Ba leaching behavior in alkaline solution. To investigate the dynamic formation of the highly active surface, we prepared epitaxial 3C BRO thin films grown on (001)-oriented Nb(0.5wt%)-doped $SrTiO_3$ (NSTO) substrates using pulsed laser epitaxy (PLE) [46]. Through repeated HER cycling using cyclic voltammetry (CV) under alkaline conditions, we evaluated the evolution of the HER activity of the

BRO catalyst. The HER activity of BRO increases substantially after the first HER cycle, gradually decreases over several subsequent cycles, and eventually saturates to an activity comparable to that of $RuO_2$ [47]. We experimentally characterized the lattice, chemical, and electronic structures of the BRO catalysts and theoretically modeled the HER activity using density functional theory (DFT) calculations. Based on the results, we propose the mechanism for dynamic surface reconstructions including formation of $Ru_6$ clusters under alkaline HER condition, as shown in Fig. 1a. During the first HER cycle, a highly active $Ru_6$/BRO structure is formed. This highly active surface gradually becomes Ru-covered with subsequent Ba leaching. Meanwhile, the highly active local BRO surface phase can be stabilized when the film is thicker, as a thick film provides BRO support to maintain the necessary balanced Ru coverage. The deduced intrinsic mechanism highlights the importance of the dynamic formation of a highly active surface for HER.

## 2. Experimental

*2.1. Thin film growth and lattice structural characterization*

The BRO thin films were grown on atomically flat, single-crystalline NSTO substrates via PLE at 500 °C. An excimer laser (248 nm, Lightmachinery, IPEX 864) with a laser fluence of ~1.0 J cm$^{-2}$ and a repetition rate of 5 Hz was used. The oxygen partial pressure was maintained at 30 mTorr during film growth. All the thin films were structurally characterized using XRD (Malvern Panalytical, X`Pert MRD) and STEM (JEOL, JEM-ARM200F).

*2.2. Chemical properties*

The core-level spectra of the BRO surfaces with/without the electrochemical cycling and further chemical stoichiometry of the BRO thin films were studied at room temperature using an XPS system (HP-XPS, K-Alpha+, Thermo Fisher Scientific, UK) with a monochromated Al-Kα X-ray source (hν = 1486.6 eV) with 400 μm spot size in diameter with charge compensation using two flood gun (low energy electron and Ar+ ion) at Yeongnam Regional Center of Korea Basic Science Institute(KBSI). For the depth profiling, we used the $Ar^+$ ion as the etching source, and the used etching energy and area were 1 kV and 1.5*1.5 $mm^2$, respectively. XAS measurements at the Ru *L*-edge were conducted at the 6A beamline of Pohang Light Source-II (PLS-II) in Pohang, Republic of Korea. Spectra were collected in total electron yield mode at room temperature with a 45° incident beam angle. Additionally, we performed STEM-EDS analysis using an EDS detector (SDD type 80T) and an analysis software (AZtecTEM, Oxford).

*2.3. Electrocatalytic performances*

Electrocatalytic reaction was measured using a potentiostat (Ivium Technologies) with a three-electrode set-up comprising a 3M NaCl saturated Ag/AgCl reference electrode, Pt mesh counter electrode, and BRO thin film as the working electrode. Note that the preparation of the working electrode with thin films was conducted using the method described in our previous studies [35, 37]. The electrolyte (0.1 M KOH solutions) was prepared by mixing de-ionized water and KOH flakes (99.99%, Sigma Aldrich). All polarization curves were obtained by sweeping the voltage at a rate of 10 mV $s^{-1}$. The measured potential converted to the RHE potential ($E_{RHE}$) by equation below.

$$E_{RHE} = E + E^0_{Ag/AgCl} + 0.0591 \times pH.$$

where $E^0_{Ag/AgCl}$ is the potential of Ag/AgCl (3 M NaCl) versus normal hydrogen electrode, which is 0.195 V at 300 K.

*2.4. Theoretical calculations*

DFT calculations were performed using the Vienna ab initio simulation package (VASP). The projector augmented wave method (PAW) was used to describe the interaction between valence and core electrons [48-50]. The general gradient approximation (GGA) with the Perdew, Burke and Ernzehof exchange-correlation functional (PBE) was employed for the plane-basis wave expansion [51, 52]. The kinetic energy cut-off of 400 eV was used and the energy convergence criteria in the self-consistent field was set to $10^{-6}$ eV. All geometry structures were fully relaxed until Hellman-Feynman forces achieved a range of 0.1 eV Å$^{-1}$. Optimization of cell parameters used a Gamma-centered *k*-point grids of 1 x 1 x 1 for Brillouin zone sampling [53]. A vacuum region of 20 Å was added to the substrate surface in the direction perpendicular to avoid interactions between the periodic slabs.

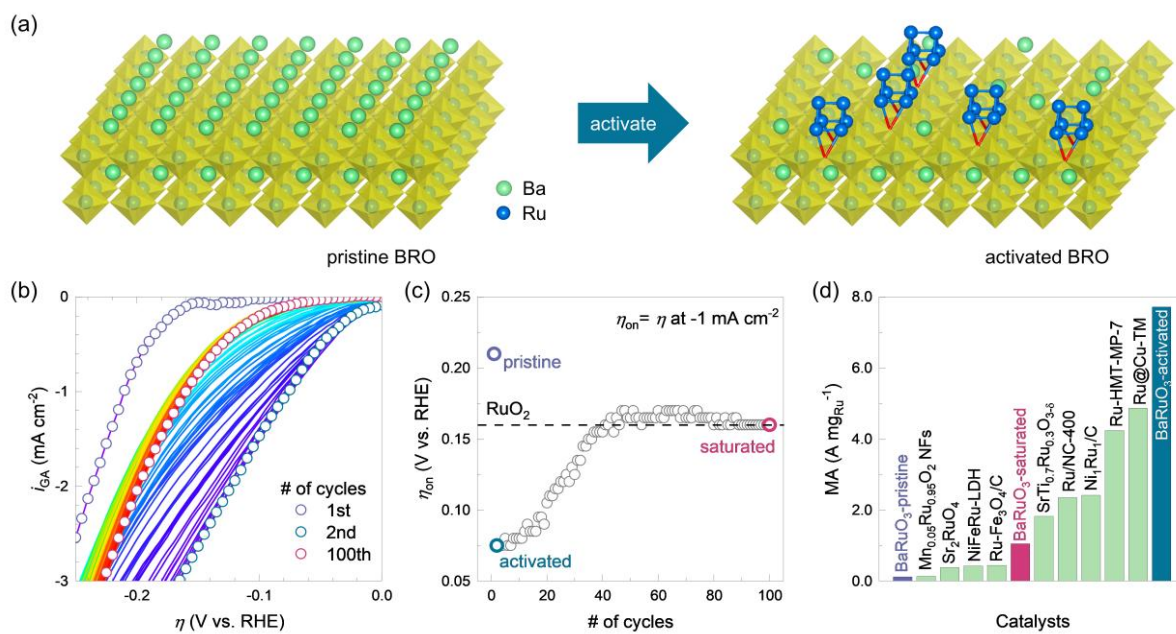

**Fig. 1.** Evolution of electrocatalytic activity with HER cycling. (a) Schematic of surface self-reconstruction. (b) CV measurement for 100 HER cycles. The 100 cycles are represented by rainbow-colored lines from violet (1st) to red (100th). The black, blue, and red squares indicate the experimental points for the 1st, 2nd, and 100$^{th}$ HER cycles, respectively. (c) HER overpotential at −1 mA cm$^{-2}$. (d) Comparison of mass activity for the Ru loading at an overpotential of 0.1 V for various Ru-based catalysts; experimental conditions for each catalyst are provided in Table S2.

## 3. Results and Discussion

We discovered that the HER activity of the BRO drastically enhanced during the first HER cycle. The HER activity of the epitaxial BRO thin film (we first focus on the 30-nm-thick film) grown on NSTO substrates was measured by CV with a three-electrode cell in the alkaline environment ($pH = 13$). The contributions of the extrinsic parameters to the HER activity were minimized by the electrochemical correction of the CV scans by the geometric surface areas (GA), electrochemical surface area (ECSA), capacitive currents estimated from the non-faradaic region, and the reversible hydrogen electrode (RHE), as described in Supplementary Note #1 and Fig. S1-S3. To quantitatively analyze the HER activity, we show the onset overpotential ($\eta_{on}$) at the $-1.0$ mA cm$_{GA}^{-2}$, which indicates the potential required for initiating hydrogen production. Surprisingly, in BRO thin film, the $\eta_{on}$ decreased significantly by as much as ~150 mV (~70%), from 210 mV to 60 mV, after the first HER cycle, which represents the highest performance reported among perovskite oxides as shown in Fig. 1b and Table S1. With increasing number of HER cycles, the activity gradually decreases and saturates after ~50 HER cycles, as shown in Fig. 1c and S4. The electrochemical phases of the BRO thin film can be classified into pristine-, activated-, and saturated-BROs, which correspond to the BRO surfaces before cycling (black), after the first HER cycle (blue), and after 100 HER cycles (red), respectively.

The activated-BRO surface exhibits significantly high intrinsic HER activity, comparable to the most active Ru-based catalyst reported under alkaline solution [54-62]. To quantitatively evaluate the intrinsic HER activity of BRO, we employed mass activity (MA) for total Ru loadings in the working electrode at a specific overpotential and Tafel slope as the activity parameters. These indicate the activity of participating active sites in the electrochemical reaction and the charge transfer coefficient according to the Tafel equation, respectively (more details in Supplementary Note #2 and Table S2). Fig. 1d shows the intrinsic HER activity of the three electrochemical phases of BRO catalyst compared to state-of-the-art Ru-based catalysts under alkaline environment. Our activated-BRO catalyst showed the highest MA value of 7.72 A mg$_{Ru}^{-1}$ at a specific overpotential of 100 mV, whereas the saturated-BRO catalyst showed 1.05 A mg$_{Ru}^{-1}$. To identify the physical and chemical origins of the activation, we have characterized the lattice and chemical structures of the three catalytic phases for the BROs.

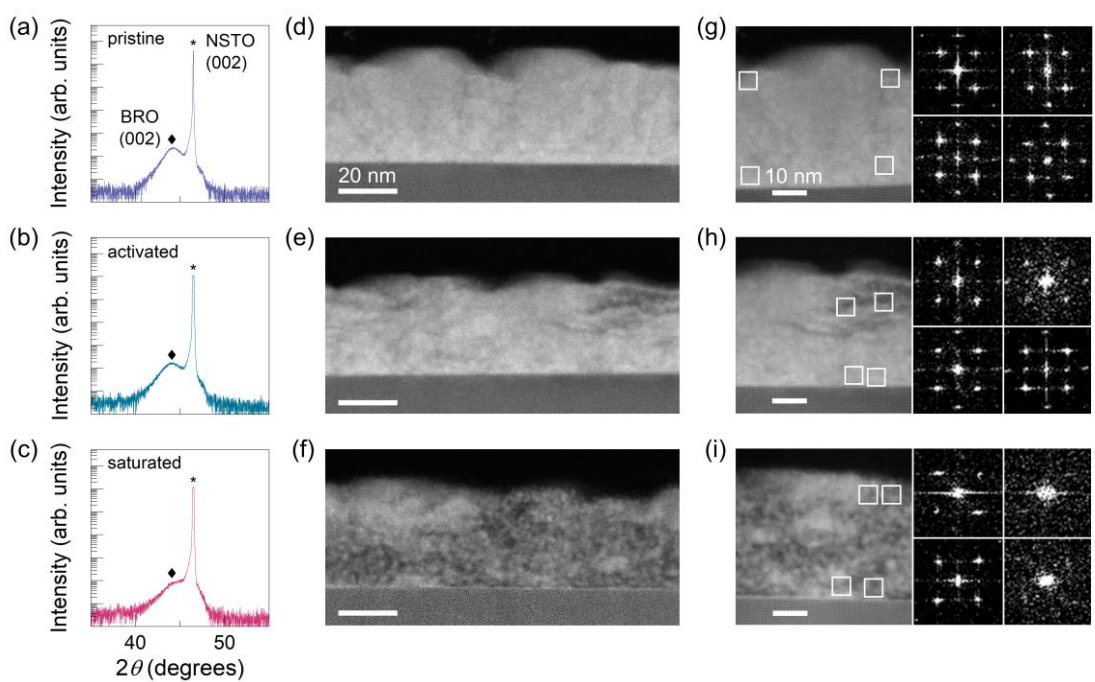

**Fig. 2**. Modification of lattice structures with HER cycling. XRD $\theta$-$2\theta$ scans near (002) Bragg peak of 3C BRO for (a) pristine, (b) activated, (c) saturated thin films. STEM images in (d–f) low- and (g–i) high-magnification of (d, g) pristine-, (e, h) activated-, and (f, i) saturated-BROs. (g–i) Fast Fourier-transforms (FFTs) of four white box regions in the STEM images are displayed on the right panels of each image.

The use of epitaxial thin film geometry enables precise tracking of the lattice structural evolution during the HER cycling. Fig. 2 shows the lattice structures of the BRO thin films characterized by X-ray diffraction (XRD) and scanning transmission electron microscopy (STEM). The pristine BRO exhibits a perovskite 3C phase, epitaxially stabilized by the cube-on-cube growth (see Fig. S5 for XRD full scan) [63]. The (002) Bragg peak intensities of the BRO thin films gradually decrease with HER cycling, indicating a gradual deformation in their crystallinity, as shown in Fig. 2a–2c. XRD rocking curves ($\omega$-scans) also support the result (see Fig. S6). STEM results also show a similar trend in crystallinity. While the pristine- and activated-BROs show 3C phase in the film region (away from the surface), the saturated-BRO shows an amorphous pattern in the FFT with diffused diffraction peaks, as shown in Fig. 2d–2i. Especially, in the case of the activated-BRO, the surface region starts to loose crystallinity after the first HER cycle. We further measured STEM-energy dispersive spectroscopy (STEM-EDS) to investigate the nature of lattice amorphization in BROs (see Fig. S7 for the STEM-EDS result). As a result, this local amorphization seems to have occurred simultaneously with the decrease in the Ba atomic concentration.

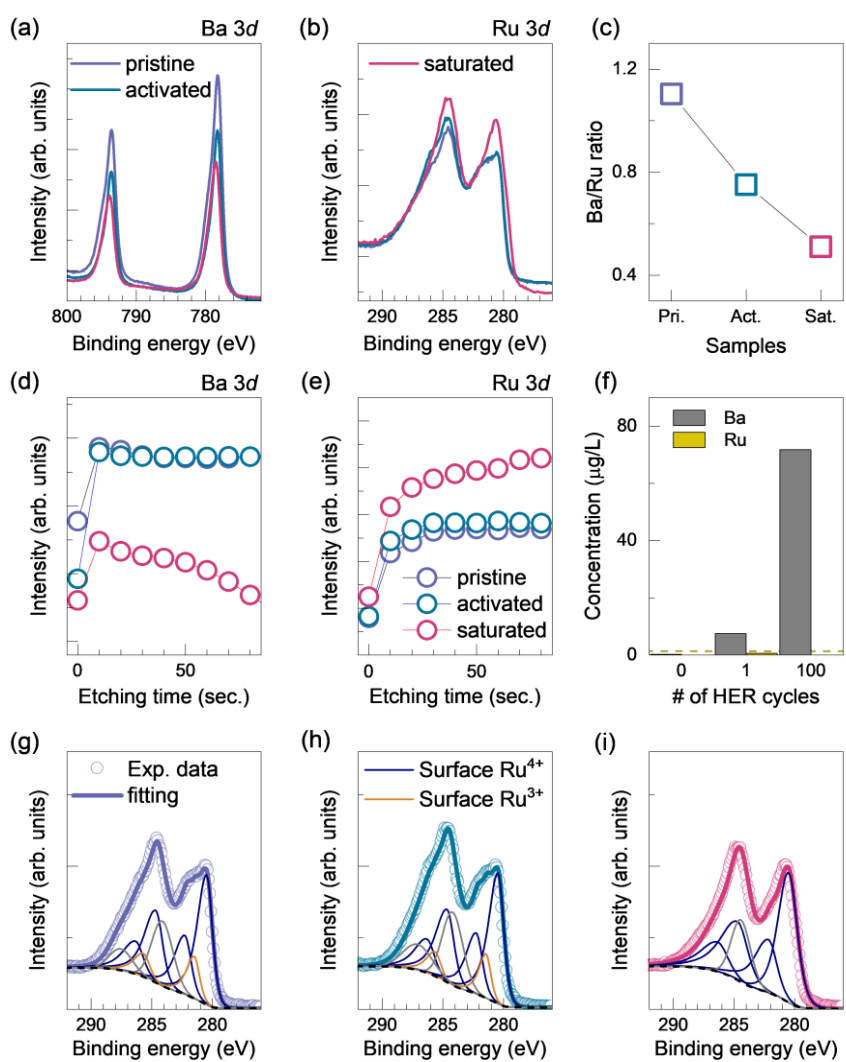

**Fig. 3.** Alteration of surface chemistry with HER cycling. XPS profiles of (a) Ba 3*d* and (b) Ru 3*d* core-level for BRO thin films with HER cycling, before Ar etching. (c) The Ba/Ru ratio determined from spectral areas of each region for Ba 3*d* and Ru 3*d* excluding carbon-related peaks. The depth-profiling results of (d) Ba 3*d* and (e) Ru 3*d* XPS spectra of BRO thin films with increasing time of Ar$^+$ ion etching. (f) Atomic concentrations of dissolved ions in electrolytes after HER cycles, in which zero-cycle means

pristine electrolyte. Note that the black and red dashed lines represent the experimental quantification limits for the Ba and Ru atoms, respectively, in the ICP-MS measurement. XPS profiles and deconvolution of peaks in Ru 3*d* spectra for (g) pristine-, (h) activated-, and (i) saturated-BRO. The navy, magenta, and grey lines indicate $Ru^{4+}$ 3*d* peaks with satellite peaks, $Ru^{3+}$ 3*d* peaks, and surface carbon-related peaks, respectively.

The structural decomposition of the BRO lattice involves the leaching of surface Ba, whereas Ru persists on the surface of the film after the HER cycles. To identify the chemical signatures of the thin film before and after the HER cycles, we conducted the X-ray photoelectron spectroscopy (XPS) measurement. From the survey spectra, we narrowed down spectral ranges for Ba 3*d*, Ru 3*d*, and O 1*s* states, as shown in Fig. 3 and Fig. S8. The survey spectra (Fig. S9) exhibit negligible signatures of unexpected peaks, confirming the growth of phase-pure BRO thin films, which is consistent with the XRD and STEM results (and no additional peaks—including Pt 4*f* at 70–76 eV—were detected in the activated and saturated samples). The most noticeable change with HER cycling is the decrease in Ba/Ru ratio of the BRO surface, as shown in Fig. 3a–3c. The surface Ba/Ru ratio decreased drastically and systematically from ~1.10 for the pristine-BRO to 0.75 and 0.51 for the activated- and saturated-BRO, respectively, indicating that the surface Ba atoms were leached out into the solution during the HER cycling. The surface Ba-deficient region in activated-BRO is diminished after a few nanometers of Ar-etching, but in the saturated-BRO, the Ba-deficient region spreads throughout the entire thin film, as shown in Fig. 3d and 3e. This is highly consistent with our STEM observation on the BRO thin films. Ba leaching from the catalyst was consistently verified by employing an inductively coupled plasma-mass spectrometer (ICP-MS) and an optical emission spectrometer (ICP-OES) of the alkaline solution after electrochemical experiments (first and 100th HER cycles), as shown in Fig. 3f and Table S3. While the pristine solution contained negligible dissolved Ba and Ru atoms, the atomic concentration of dissolved Ba atoms significantly increased with the HER cycling. Meanwhile, the Ru atoms in the BRO exhibited robust stability under alkaline HER conditions. Despite local amorphization and Ba leaching shown in the saturated-BRO, the dissolution of Ru into the solution remained below the measurement limit even after the one hundred HER cycles.

It is evident that the Ru ions play a decisive role as the electrochemically active site for the HER of the BRO catalyst. Especially, the oxidation state of Ru may influence the activity. For example, adding $Ru^{3+}$ atoms on mesoporous carbon support was reported to be beneficial for achieving an efficient and durable HER [64]. Thus, we identified the surface valence state of Ru from the Ru 3$d$ XPS spectra, as shown in Fig. 3g–3i (before Ar etching). Note that the fitting results of our BRO thin films are consistent with previous literatures [37, 65-68], with further details provided in Supplementary Note #3. From the XPS results, unlike the saturated-BRO, the pristine- and activated-BROs exhibit additional peaks associated with the low-oxidation state of Ru ions, which could originate from the oxygen vacancy or formation of a local Ru-rich phases as shown in Fig. 3g-3i. The $Ru^{3+}$ states eventually disappear in the saturated-BRO surface, with the saturation of the HER activity. The low-oxidation Ru sites on the surface disappear after 30 seconds of Ar etching, as shown in Fig. S10, suggesting that they are localized near the surfaces (approximately 6 nm from the surface) of the thin films. Importantly, the $Ru^{3+}/Ru_{total}$ ratio remains essentially unchanged (Fig S11), for which the surface Ba/Ru ratio changes drastically between the pristine and activated thin films. The X-ray absorption spectroscopy (XAS) demonstrates negligible changes in the average Ru oxidation state of the thin film region with increasing HER cycles, as shown in Fig. S12, supporting that the Ru sites with low oxidation states are confined near the surface region in both pristine- and activated-BRO catalysts. The XPS and XAS on low-oxidation state of Ru ions imply that the HER activity is mainly governed by the Ba leaching, but not by the presence of $Ru^{3+}$. To directly identify Ru rich surface, we employed high-resolution STEM on pristine BRO and BRO in the highly active phase (after 10 HER cycles), as shown in Fig. S13. Although it would be highly challenging to observe the exact nanoclusters using STEM [69], it exhibits a Ru-rich surface with possible Ru clusters following the HER cycling, consistent with the STEM-EDS results for the activated BRO, as shown in Fig. S6, and X-ray spectroscopies discussed above.

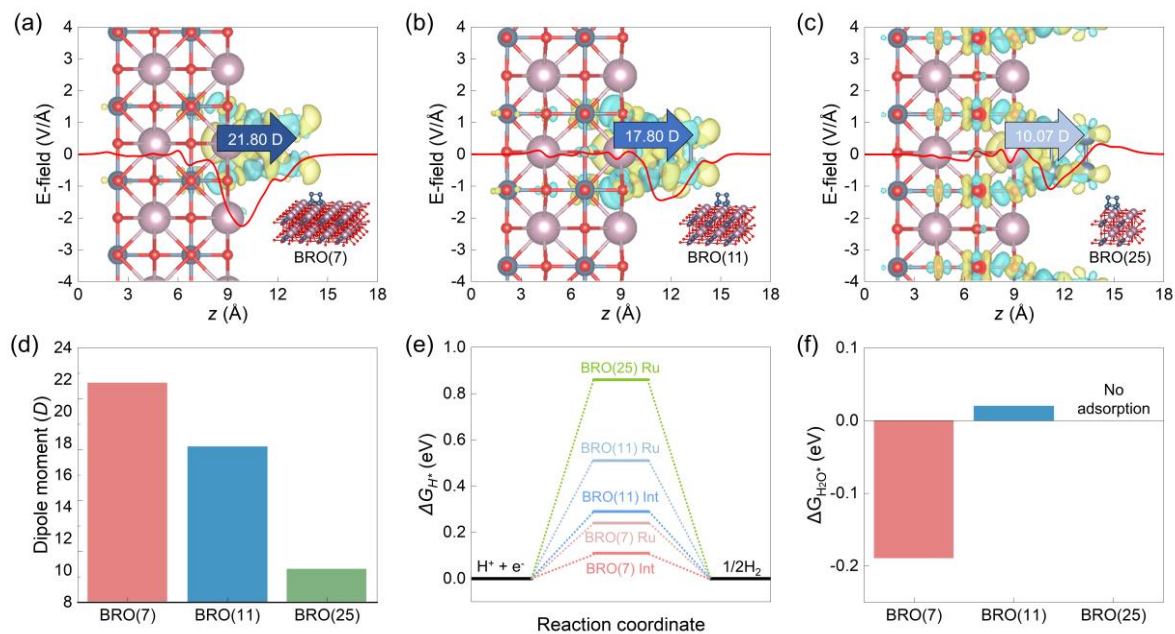

**Fig. 4.** DFT calculation results of the built-in electric field and the modeled supercells of (a) BRO(7), (b) BRO(11), and (c) BRO(25) models. (d) The magnitude of dipole moment at $Ru_6$/BRO surface with different Ru coverages. (e) Gibbs free energy diagram for the H* intermediate with the reaction coordinate and (f) Gibbs free energy profiles for the $H_2O$* intermediate in $Ru_6$/BRO surfaces.

The formation of surface-localized Ru clusters on the BRO surface is thermodynamically favorable under the experimental conditions for alkaline HER. Motivated by the experimentally observed surface Ba leaching, we conducted DFT calculations to understand the formation of $Ru_6$/BRO structure during HER cycling and its impact on the electrocatalytic activity (Fig. 4). It has been reported that the $Ru_6$ configuration is thermodynamically favored under alkaline HER conditions among various Ru clusters [30]. Consistent with previous study, the surface Gibbs free energy ($\Delta G^S$) for $Ru_6$ formation on BRO at pH 13 remain ≤ 0 across the experimental bias window (Fig. S14), indicating that $Ru_6$ formation is thermodynamically accessible under our conditions. Therefore, the $Ru_6$ cluster was selected as the primary phase of the Ru cluster on the BRO surface. To investigate the preferred Ru-rich surface configuration on the BRO surface, we modeled three supercells with Ru surface coverages of 7%, 11% and 25%, denoted as BRO(7), BRO(11), and BRO(25), respectively. The schematic representations of the supercells are shown in the insets of Fig. 4a–4c and Fig. S15. Across the experimental bias range, the formation of Ru$_6$ clusters is energetically favored and expected to become more abundant with time, suggesting that the Ru$_6$ coverage would increase with electrocatalytic cycling.

Reconstructed BRO surfaces with varying Ru coverages induce electric dipole moments, adjusting the built-in electric field and resultant electrostatic potential barrier for the adsorbate molecules. When BRO and the Ru$_6$ cluster form a heterojunction, the charges near the interface are redistributed, which can be represented by the planar-averaged charge density along the $z$-axis, as shown in Fig. S16. The inward built-in electric field and resultant electric dipole moment at the Ru$_6$/BRO surface are determined by the charge density distribution, as shown in Fig. 4a–4c. As the coverage of Ru$_6$ cluster decreases, more electrons transfer from the BRO surface to the Ru$_6$ cluster, leading to an increase in the magnitude of the dipole moment ($D$), as shown in Fig. 4d. Note that the $D$ can either enhance or hinder the adsorption of adsorbates [70], from which the electric potential ($\Delta V$) and electric field ($\vec{E}$) can be defined with respect to the adsorption sites on Ru$_6$ cluster, as shown in Supplementary Note #4. Under alkaline conditions, the HER process involves the activation of H$_2$O, producing adsorbed H* species and OH$^-$ ions in the electrolyte. For OH$^-$ ions, the inward built-in electric field can act as a repulsive force against negatively charged species, thereby driving the generated OH$^-$ ions away from the surface. As a result, the H* coverage is enhanced, therefore improving the HER performance under basic conditions.

The Gibbs free energy of HER adsorbate ($\Delta G_{ads^*}$) on the Ru$_6$/BRO surface can be rigorously determined by incorporating electrostatic potential energy terms induced by the interface polarization as follows [70],

$$\Delta G_{ads^*} = \Delta E_{ads} + \Delta ZPE + \int C_p dT - T\Delta S + \Delta G_{pH} + e\Delta V \quad \text{(Eq. 1)}$$

where $\Delta E_{ads}$, $\Delta ZPE$, $\int C_p dT$, and $\Delta S$ are the adsorption energy, the change of zero-point energy, the enthalpy, and the change of entropy, respectively. $\Delta ZPE$, $\int C_p dT$, and $T\Delta S$ can be written as,

$$\Delta ZPE = \frac{1}{2}\sum \hbar\omega_i \quad (Eq.\ 2)$$

$$\int C_p dT = \sum_i \frac{\hbar\omega_i}{exp\left(\frac{\hbar\omega_i}{k_B T}\right) - 1} \quad (Eq.\ 3)$$

$$T\Delta S = \sum_i \frac{\hbar\omega_i}{\exp\left(\frac{\hbar\omega_i}{k_B T}\right) - 1} - k_B T \sum_i \ln\left(1 - \exp\left(\frac{\hbar\omega_i}{k_B T}\right)\right) \quad (Eq.\ 4)$$

where $\hbar$, $\omega_i$, and $k_B$ are the reduced Planck constant, the vibrational frequency eigenvalue of $i$-th mode, and the Boltzmann constant, respectively. The $\Delta G_{pH}$ term in Eq. 1 is included to consider the $pH$ effect on the reactivity, as $k_B T ln10 \times pH$. The $\Delta V$ was included in Eq. 1 (i) the induced electrostatic potential by the polarization at $Ru_6$/BRO surface and (ii) the difference of electrostatic potential between the catalyst surface and hydrogen reduction level into account. Regarding the potential active sites of $Ru_6$/BRO, we considered Ru top site (BRO Ru) and the site near the interface (BRO Int), as indicated in Fig. S17. Additionally, an adsorption site far from the $Ru_6$ cluster was examined on BRO(7) (BRO Far) only, as more Ba atoms are exposed on the surface of BRO(7).

The DFT calculation indicates that the H* adsorption on the BRO surface only becomes possible as $Ru_6$ clusters are formed. In other words, $Ru_6$/BRO structure promotes H* adsorption, whereas the pristine BRO surface prevents it (Fig. S18). In alkaline HER, the hydrogen adsorption energy ($\Delta G_{H^*}$) is one of the key factors influencing the activity owing to the essential role of the H* intermediate in the reaction mechanism. [71]. Fig. 4e and Table S4 show the calculated H* adsorption energy on different BRO surfaces. The differences in H* adsorption from the calculation explain the low performance of pristine BRO surface compared to the activated BRO surfaces observed in the experiments. According to Sabatier's principle, the balance of H* adsorption and desorption with $\Delta G_{H^*}$ being close to zero is important for the HER performance [72]. The calculated H* adsorption energy of BRO(7) Int is the closest to zero (0.11 eV), whereas that of BRO(25) Ru is 0.86 eV. It reveals that the high coverage of $Ru_6$ clusters on BRO surface can degrade the catalytic activity of HER, which also aligns well with the experimental results.

As Ru coverage further increases, H$_2$O adsorption is hindered by electrostatic repulsion at the Ru$_6$/BRO surface, decreasing the HER activity. The activation of H$_2$O in the Volmer step is another crucial reaction process in HER under alkaline condition [73]. Therefore, we also calculated adsorption energy of H$_2$O intermediate for the Ru$_6$/BRO surfaces of BRO(7), BRO(11), and BRO(25), as shown in Fig. 4f and Table S4. Fig. S19 schematically represents how a H$_2$O molecule is adsorbed on BRO Int sites of BRO(7) and BRO(11), while the adsorption of H$_2$O on BRO Ru sites did not occur. Note that H$_2$O molecules were not bound to any adsorption sites of BRO(25). Based on the calculation, H$_2$O activation is thermodynamically more favored at the Ru$_6$/BRO surfaces of BRO(7) compared to the other Ru coverages. Consequently, the BRO(7) Int site effectively facilitates the H$_2$O adsorption, ensuring a sufficient supply of reactant molecules for alkaline HER. However, as Ru coverage increases, the availability of H$_2$O adsorption sites diminishes, potentially limiting the overall reaction efficiency and thereby decreasing HER activity, as observed in the experiment.

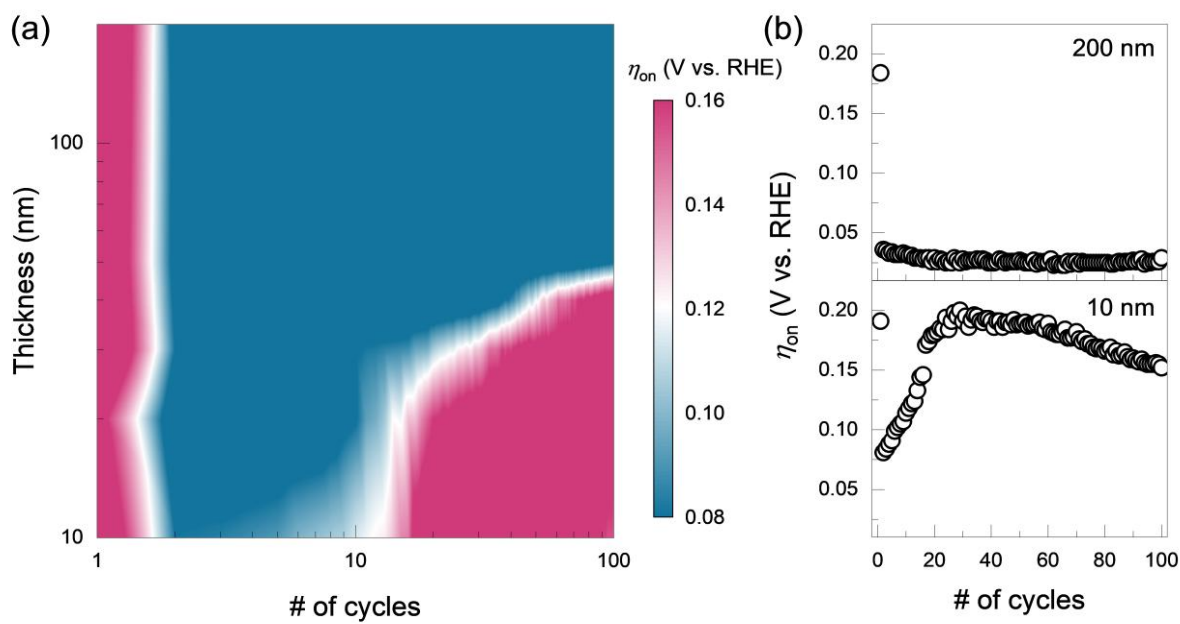

**Fig. 5.** Stabilization of the active phase of the BRO catalyst through film thickness engineering. (a) A color map representing the onset overpotential as a function of the number of HER cycles and film

thickness. (b) Variation of onset overpotential trends for BRO thin films with thickness of 10 (top panel) and 200 nm (bottom panel).

Based on these experimental and theoretical observations, we propose a step-by-step mechanism to explain the HER activity trend of the 30-nm-thick BRO epitaxial film as follows. i) During the activation process of the first HER cycle, leaching of surface Ba atoms into the alkaline solution occurs. The pristine BRO surface transitions into a Ru-rich state, forming a local Ru-cluster-covered surface. The reconstructed surface facilitates the formation of $Ru_6$ clusters, which are thermodynamically stabilized by interactions with the BRO support layer. We note that such Ru clusters may form on other perovskite Ruthenates such as $SrRuO_3$, but Sr leaching was found to occur not as dramatic as Ba leaching (Fig. S20). ii) In high-activity cycle region (2–20 HER cycles), the highly active $Ru_6$/BRO structure with relatively low Ru coverage is sustained by the continuous migration of Ba atoms from the BRO support layer to the surface, maintaining optimal catalytic phase for alkaline HER. iii) In the activity-decreasing region (20–50 HER cycles), Ba is depleted due to the continued Ba leaching, and Ru surface coverage starts to increase. This reduces the HER activity. iv) In the saturated region (>50 HER cycles), a relatively high Ru surface coverage is realized. Excessive Ru coverage drives the BRO surface state similar to that of $RuO_2$, leading to HER activity comparable to that of rutile-$RuO_2$ catalyst.

To effectively utilize our discovery, we propose a practical strategy extending the activated phase of the BRO surface employing a thicker film. The suggested mechanism implies that maintaining a highly active surface requires the presence of the prevailing BRO sub-surface. Hence, we demonstrated the film thickness-dependent HER activity of the BRO epitaxial thin films under alkaline solutions, as shown in Fig. 5a. When the film is thin (~10 nm), the high activity corresponding to the activated-BRO surface disappears quickly within 15 HER cycles (lower panel of Fig. 5b). On the other hand, when the film thickness is increased to about 50 nm, the high activity after the first HER cycle was maintained even after 100 HER cycles. As shown in the upper panel of Fig. 5b, the 200-nm-thick BRO film does not show any signs of degradation of the activity up to 100 HER cycles, manifesting the dramatic enhancement of the stability of the catalytically awakened surface in the thicker films. Consistently, chronoamperometry at a fixed overpotential shows the same trend, with the 200 nm film maintaining high activity for a longer duration than the 30 nm film (Fig. S21). Especially, we note that the stability of the catalytically active surface does not increase linearly with the film thickness. Above 50 nm, the electrocatalytic stability of the awakened active surface is drastically improved with increasing thickness, e.g., the highly active phase was maintained even after 1000 HER cycles for a 800-nm-thick film (Fig. S22). These thickness-dependent trends indicate that a sufficiently thick BRO support layer buffers Ba leaching and preserves low Ru coverage surface under bias, thereby maintaining the awakened highly active phase. Because powder or self-supported electrodes typically possess a much larger thickness than our thin-film system, this stabilization principle is readily transferable—guiding the design of scalable, durable HER catalysts for device-relevant operation.

## 4. Conclusion

A mechanism for awakening a highly active BRO catalytic surface for the HER by introducing a $Ru_6$/BRO structure was discovered. The HER of the BRO catalyst was significantly activated by single-voltage cycling in an alkaline solution, promoting high HER activity comparable to that reported for the most active ruthenate catalysts. As the BRO exhibited high HER activity, leaching of Ba atoms near the surface was observed. However, increasing the Ru coverage on the BRO surface beyond a threshold decreased the HER activity. Through DFT calculations, the introduction of $Ru_6$ clusters on the surface of BRO was identified, which enhanced HER activity by promoting H* adsorption. The calculations revealed that as the Ru coverage increased, the HER activity decreased and eventually saturated. We further propose a strategy to dynamically stabilize the catalytically awakened surface by film thickness engineering. Beyond thin films, the role of the support layer can be applied to powders or supported architectures (e.g., core–shell or oxide-scaffold systems). Here, the 'effective thickness' may be controlled to stabilize low Ru coverage during cycling and thus improve durability in practical HER catalysts. Our study provides fundamental insights into the formation of a highly active and stable BRO catalyst surface, offering valuable guidance for designing high-performance electrocatalysts for the alkaline HER.

## Acknowledgements

This work was supported by the National Research Foundation of Korea (NRF) funded by the Ministry of Science and ICT (NRF-2021R1A2C2011340, NRF-RS-2023-00220471, NRF-RS-2023-00281671). This research was supported by Basic Science Research Program through the NRF funded by the Ministry of Education (NRF-2022R1I1A1A01065234). H.C. Acknowledges the supports from the Suzhou Science and Technology Development Planning Programme (SYC2022101), the Suzhou Industrial Park High Quality Innovation Platform of Functional Molecular Materials and Devices (YZCXPT2023105), the XJTLU Advanced Materials Research Center (AMRC). J.S.B. was supported by the Synchrotron Strategic Materials Analysis Research Program (C526111) funded by the Korea Basic Science Institute.